\documentclass{elsart}
\usepackage{amsfonts}

\usepackage{graphicx}
 \usepackage{epsfig}
 \usepackage{subfig}
\usepackage{amssymb}
\usepackage{amsmath}

\begin{document}

\begin{frontmatter}

\title{\large{\bf Landing together: how flocks arrive at a coherent action in time and space in the presence of perturbations }}

\author{B. Ferdinandy$^{a,*}$, K. Bhattacharya$^{a,b}$, D. \'Abel$^{a}$, T. Vicsek $^{a,c}$}

\corauth[1]{Corresponding author. E-mail:
fbence@elte.hu}
\address
{$^a$Department of Biological Physics, E\"otv\"os University, Budapest,
 H-1117 Hungary\\
$^b$Department of Physics, Birla Institute of Technology and Science, Pilani 333031, India\\
$^c$Statistical and Biological Physics Group, ELTE-HAS, 
P\'azm\'any P. Stny. 1A, Budapest, H-1117 Hungary\\
}

\begin{abstract}
Collective motion is abundant in nature, producing a vast amount of phenomena which have been studied in recent years, including the landing of flocks of birds. We investigate the collective decision making scenario where a flock of birds decides the optimal time of landing in the absence of a global leader. We introduce a simple phenomenological model in the spirit of the statistical mechanics-based self-propelled particles (SPP-s) approach to interpret this process. We expect that our model is applicable to a larger class of spatiotemporal decision making situations than just the landing of flocks (which process is used as a paradigmatic case). In the model birds are only influenced by observable variables, like position and velocity. Heterogeneity is introduced in the flock in terms of a depletion time after which a bird feels increasing bias to move towards the ground. Our model demonstrates a possible mechanism by which animals in a large group can arrive at an egalitarian decision about the time of switching from one activity to another in the absence of a leader. In particular, we show the existence of a paradoxical effect where noise enhances the coherence of the landing process.
\end{abstract}

\end{frontmatter}

\section{Introduction}

Scenarios involving collective decision making appear to be ubiquitous in several fields including animal behaviour \cite{conradt-1} and social
sciences \cite{chamley}. Several studies \cite{garnier,ward} have investigated
the cases where any group unanimously decides to choose one of the many available
options. It is generally expected that there will be differences in the motivations of the members at the time when making choices.  However, in spite of the differences consensus is seen to evolve. Different mechanisms have been suggested to account for this type of process \cite{detrain,bouchaud,castellano}. The related phenomenon of collective motion in animal groups has
also been extensively modelled \cite{couzin,gregoire,vicsek} by
considering animals as point-like particles (so-called self-propelled particles). In these models it is
assumed that each individual moves with constant speed while tending to align with its immediate
neighbours, for low level noise, giving rise to a globally ordered state replicating the motion of flocks
where animals move in the same direction.

Our object of study is the phenomenon during which the animals moving in groups seemingly make unanimous decisions on the time and choice of performing activities \cite{conradt-1,conradt-3} even in the absence of global leaders. Examples include takeoff of swarm of honeybees from nest sites \cite{visscher}, activity synchronisation in Merino sheep \cite{gautrais}, collective movement of white-faced capuchins \cite{meunier}, group departures of domestic geese \cite{ramseyer} and departure of Argentine ants from feeding site \cite{halley}. Such a paradigm could be generalised to human behaviour as well, e.g. where to stop to rest, when making an excursion with a group. The important extra feature of taking into account group
motion during collective decision making is that in such an approach the neighbourhood with which a
consensus is to be achieved is dynamically changing in a realistic manner.

In this report we model the process of landing of bird flocks performing foraging flights as a typical
example of collective decision making. We regard the birds as self-propelled particles, the only
difference between the birds being an a priori value corresponding to the heterogeneity in
motivations. Throughout this paper we shall use the terms ``birds''
and ``particles'' interchangeably. In a recent model for collective landing \cite{daruka} the birds are assumed to move
under the action of different social forces \cite{helbing}. In addition, the internal state
of each bird is characterised by a continuous variable called landing intent such that the internal
state of each bird is directly coupled to the internal state of its neighbours. Another model \cite{bhattacharya} allows the motivation of individual birds to be influenced by only the
observable variables of their neighbours, like velocity and position. This assumption simplifies the
description of the landing process, since it does not presume very sophisticated channels of
communication during flight. The model discussed in this report has also been formulated along
similar lines. The decision of a bird is influenced by the state of motion of its neighbours and an a
priori intent (a pre-assigned constant) taking into account its endurance that acts through a noise variable. This makes the landing a
stochastic process even if we consider a single bird only.

Our aim with this report is to quantitatively show that a group without leaders
can perform a synchronised landing following very simple rules, even when a single bird does not
have information about the whole of the flock. One of the important new features of our approach is
the possibility of tuning the level of perturbations that the particles are subject to during the landing
process. This approach also allowed us to uncover an interesting effect. According to our simple model,
there is an optimal level of perturbations resulting in the smallest spread of the landing times of the
individual birds, meaning that in certain cases the noise can enhance the coherence of the landing process.

\section{Model}

\subsection{Basics}

In this section, we will describe the basic concepts of our model, leaving the majority of the technical
details to the next section. We formulate the dynamics of a bird flock along the lines of social forces. These forces are similar, but not the equivalent of forces
considered in earlier studies \cite{couzin,aoki,reynolds}. In the model, we
consider a flock that performs a horizontal flight about an average height until it decides to land. This
allows us to handle the horizontal and vertical motion of the birds separately.

Let us first regard the horizontal motion. Setting $z=0$ as the only natural boundary, representing the ground, the horizontal flight is performed parallel to the $\textrm{xy}$-plane. There are four social forces acting upon the $i$th bird horizontally, (i) $\mathbf{f}_{\textrm{xy}}^\textrm{a}$ - an averaging on the velocities of other birds in its neighbourhood, (ii) $\mathbf{f}_{\textrm{xy}}^\textrm{n}$ - an evenly distributed noise i.e., a random perturbation on its velocity, (iii) $\mathbf{f}_{\textrm{xy}}^\textrm{r}$  - a repulsive force, and (iv) $\mathbf{f}_{\textrm{xy}}^\textrm{c}$ - a cohesive force. The force $\mathbf{f}_{\textrm{xy}}^\textrm{a}$ is responsible for aligning the velocities of neighbouring birds in the flock. For this purpose we define the neighbourhood ($\mathcal{N}_i$) of the $i$th bird as those birds that are encompassed by a cylinder of infinite height and radius $R$ centred on the $i$th bird. While in flight a bird would like to maintain some separation with other birds. This is accounted by the repulsive force. This force is characterised by a certain radius of interaction $d$, which can also be considered as the effective ``size of a bird'' and is much lower than the interaction radius characterising $\mathcal{N}_i$. The purpose of the cohesive force is to keep the flock from breaking apart. It is described by another infinite cylinder with a radius $D$ which is much larger than that of $\mathcal{N}_i$, and is centred on the centre of mass ({\it CoM}) of all $N$ birds. If a bird strays out of this cylinder it is propelled towards the centre of mass.

The vertical motion also consists of four social forces: (i) $f_\textrm{z}^\textrm{a}$ - an averaging on the vertical velocities of the neighbourhood $\mathcal{N}_i$  of the $i$th bird, (ii) $f_\textrm{z}^\textrm{h}$ - a force that tries to keep the birds about a given height, (iii) $f_\textrm{z}^\textrm{r}$ - a repulsive force like in the vertical direction, and (iv) $f_\textrm{z}^\textrm{n}$  - a noise. The force $f_\textrm{z}^\textrm{h}$ is such, that around a given height $h$ there is a regime of width $\Delta h$ where in all practical sense, a bird moves freely, but outside of that, the bird is repelled toward $h$. The most important part of the model is the vertical noise, since this is the force that facilitates the landing of the birds. This is done as follows: each bird is assigned an a priori value, a $t_i$ depletion time, which represents the time at which the bird starts to feel the depletion of its energy reserves, and would increasingly wish to land. In general these depletion times will depend on several external or internal conditions such as the energy reserves of the birds \cite{sirot}, stamina, health, willingness to fly and other things and thus will be in general different for different birds. As such, these values are drawn from a Gaussian distribution of a mean $\mu$ and a variance $\sigma^2$. When a bird $i$ reaches its $t_i$, the originally evenly distributed $f_\textrm{z}^\textrm{n}$ noise starts to get increasingly biased to facilitate landing with a characteristic time $\tau$. Landing occurs when this force and the averaging force overwhelms $f_\textrm{z}^\textrm{h}$, allowing the birds to land. Thus the equations of motion of the $i$th bird in the horizontal direction are:
\begin{equation}
\mathbf{f}_{\textrm{xy},i}^{\textrm{sum}}= \mathbf{f}_{\textrm{xy},i}^\textrm{a} + \mathbf{f}_{\textrm{xy},i}^\textrm{n} + \mathbf{f}_{\textrm{xy},i}^\textrm{c}+\mathbf{f}_{\textrm{xy},i}^\textrm{r}
\end{equation}
and
\begin{equation}
\mathbf{r}_{\textrm{xy},i}(t+\Delta t)=\mathbf{r}_{\textrm{xy},i}(t)+v\frac{\mathbf{f}_{\textrm{xy},i}^{\textrm{sum}}}{|\mathbf{f}_{xy,i}^{\textrm{sum}}|}\Delta t,
\end{equation}

where $\mathbf{r}_{\textrm{xy},i}$ is the position of the $i$th bird in the $\textrm{xy}$-plane, and $v$ is a constant, so that the birds move with a velocity of constant magnitude. Very similar equations hold for the vertical motion:
\begin{equation}
f_{\textrm{z},i}^\textrm{sum} = f_{\textrm{z},i}^\textrm{a}+f_{\textrm{z},i}^\textrm{n}+f_{\textrm{z},i}^\textrm{h} + f_{\textrm{z},i}^\textrm{r}
\end{equation}
\begin{equation}
z_i(t+\Delta t)=z_{i}(t)+v\frac{f_{\textrm{z},i}^{\textrm{sum}}}{|f_{\textrm{z},i}^{\textrm{sum}}|}\Delta t,
\end{equation}

where $z_i$ is the vertical position of the bird. As seen in the next subsection, the magnitude of these forces are parameters of the model, which seemingly are a lot in number, but only some of them are actually relevant to the problem of landing. Therefore we study the behaviour of the system as a function of a few parameters, {\it viz}. the variance of depletion times $\sigma$, the number of birds $N$, and the coefficients of  $f_{\textrm{z},i}^\textrm{h}$ and $f_{\textrm{z},i}^\textrm{n}$ (described in \ref{details}). The characteristic time $\tau$ is only important in the sense that it must be of the order of magnitude of $\sigma$ for non-trivial behaviour of the model. In addition, all the differences between the birds are contained in $\sigma$.  

\subsection{Details}
\label{details}

\begin{figure}
\begin{center}
\subfloat[]{\includegraphics[scale=0.27, totalheight = 5 cm]{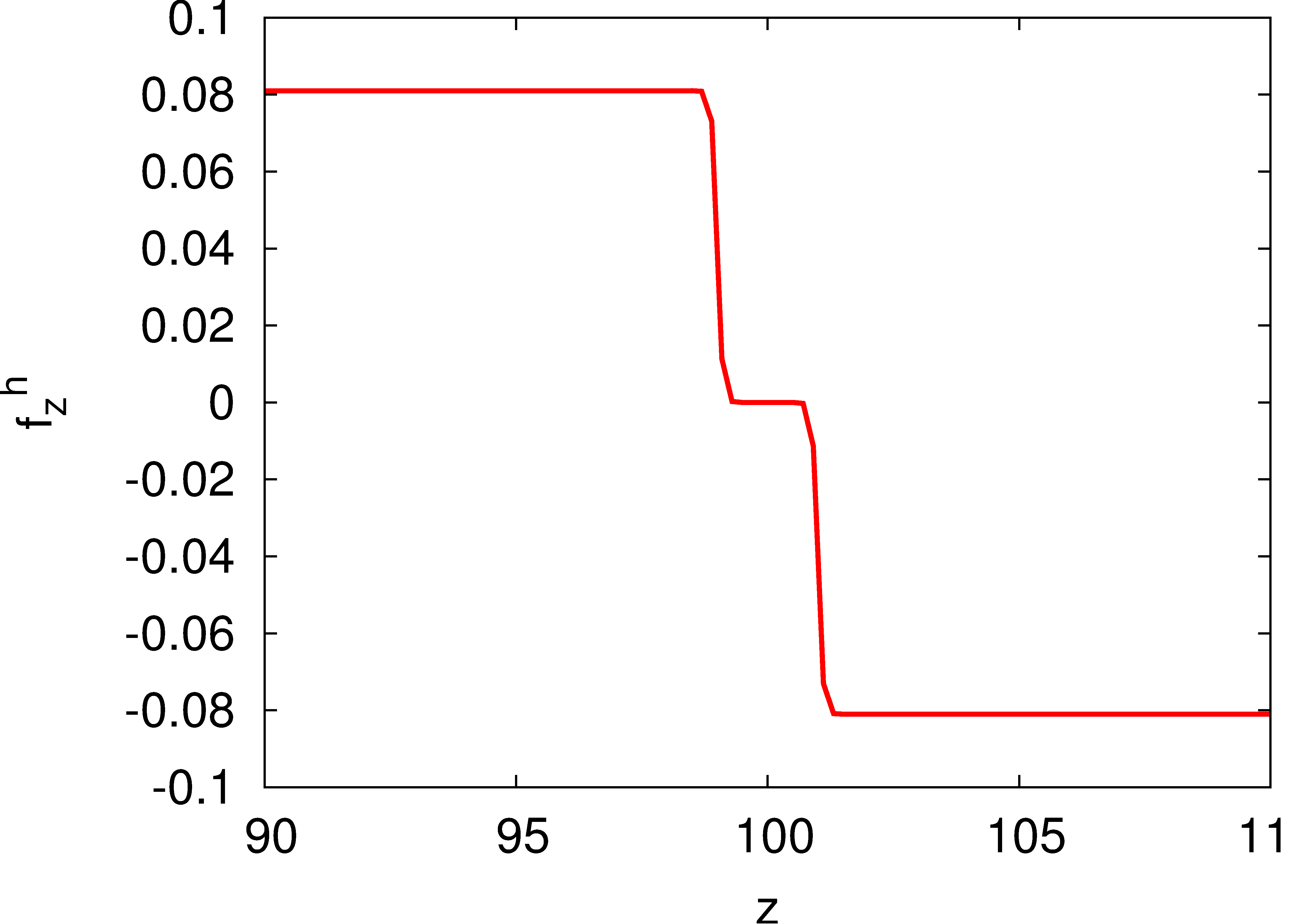}\label{fig:fig1sub}}
\subfloat[]{\includegraphics[scale=0.27, totalheight = 5 cm]{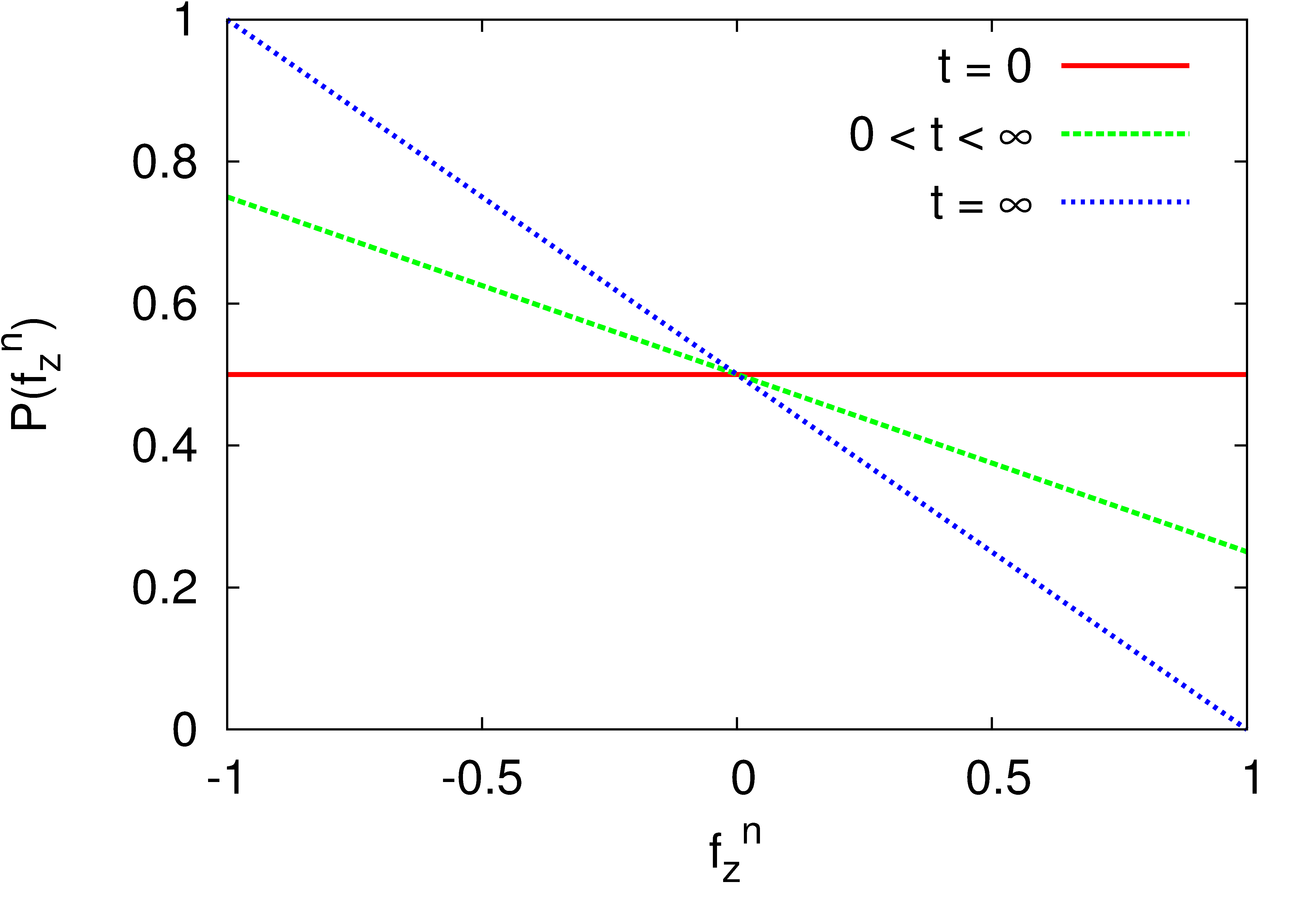}\label{fig:fig2sub}}
\end{center}
\caption{(a) The force $f_\textrm{z}^\textrm{h}$ against $z$. The plot shows the small forceless regime around $h$ and the fast strengthening of the force
outside of that, quickly saturating to a constant. (b) The plot shows the time evolution of the probability distribution of $f_\textrm{z}^\textrm{n}$. Equation \ref{eq-13} is
the explicit formula for generating values of this force.
}
\label{forcefig}
\end{figure}

In this section we provide the exact mathematical expressions for the forces, and other details concerning the model. Noting that $\mathbf{v}_{\textrm{xy}}$ and $v_{\textrm{z}}$ stand for the instantaneous velocities in the horizontal plane and the vertical direction respectively, the averaging forces are given by: 
\begin{equation}
\mathbf{f}_{\textrm{xy},i}^\textrm{a}  =  \langle\mathbf{v}_{\textrm{xy}}{\rangle}_{\mathcal{N}_i}
\end{equation}
\begin{equation}
f_{\textrm{z},i}^\textrm{a} = \langle v_{\textrm{z}}{\rangle}_{\mathcal{N}_i}
\end{equation}
where $\langle\phantom{z}{\rangle}_{\mathcal{N}_i}$ is the average over the birds in the neighbourhood $\mathcal{N}_i$ of the $i$th bird. We also assume that the averaging includes birds which have already landed (in $\mathcal{N}_i$) so that their influence is taken into account. The effect of the landed birds on the averaging forces would be the same as that of those moving vertically downwards. The repulsive forces follow the same scheme in all three dimensions. The explicit form for the $z$-direction is as follows
\begin{equation} 
f_{\textrm{z},i}^\textrm{r}= \sum_{j=1}^{N}f_{\textrm{z},ij}^\textrm{r}
\end{equation}
where
\begin{equation}
f_{\textrm{z},ij}^\textrm{r}=\left\{\begin{array}{ll}
A\left(d-|z_i-z_j|\right) & \textrm{if}\phantom{x}0<|z_i-z_j|<d\\
\phantom{zzzzz}0 & \textrm{otherwise.}
\end{array}\right.
\end{equation}
Here $A$ is the strength of the repulsive force. The cohesive force on the $i$th bird is given by the following equation:
\begin{equation}
\mathbf{f}_{\textrm{xy},i}^{\textrm{c}}=\left\{\begin{array}{ll}
\phantom{zzzzz} 0 & \textrm{if}\phantom{x}|\mathbf{r}_{\textrm{xy},i}^{\Delta \textrm{\it CoM}}|\leq\frac{D}{2}\\
-B \left(|\mathbf{r}_{\textrm{xy},i}^{\Delta \textrm{\it CoM}}|-\frac{D}{2}\right) & \textrm{otherwise,}
\end{array}\right.
\end{equation}
where $\mathbf{r}_{\textrm{xy},i}^{\Delta \textrm{\it CoM}}$ is its distance from the {\it CoM} and $B$ is the strength of the force. The form of the force ensures that if a bird is farther from the {\it CoM} than $D/2$ it feels an attraction towards the {\it CoM}. The social force describing the will to stay at a given height $h$ is given by
\begin{equation}
f_{\textrm{z},i}^{\textrm{h}}=-\frac{C}{20}\left[1+\textrm{tanh}\left\{\frac{10}{R}\left(|z_i-h|-\frac{\Delta h}{2}\right) \right\}\right]\textrm{sign}(z_i-h),
\end{equation} 
where $C$ is the strength of the force. The attraction $f_{\textrm{z},i}^{\textrm{h}}$ towards the preferred altitude $h$ is actually weak within a region of width $\Delta h$ and is strong outside (see figure \ref{fig:fig1sub}). 

We now define the nature of the noise in the vertical and the horizontal directions. At any instant of time the $i$th bird is influenced by a horizontal noise:
\begin{equation}
\mathbf{f}_{\textrm{xy},i}^{\textrm{n}}(t)=\beta{\mathbf{\xi}}_{\textrm{xy},i}(t),
\end{equation}
where ${\mathbf{\xi}}_{\textrm{xy},i}$ is a unit vector on the $xy$-plane whose orientation is taken to be random and $\beta$ is the strength of the noise. For times less than $t_i$ the vertical noise $f_{\textrm{z},i}$ is given by:
\begin{equation}
f_{\textrm{z},i}^{\textrm{n}}(t)=\alpha\xi_{\textrm{z},i}(t),
\end{equation}
where $\xi_{\textrm{z},i}(t)$ is randomly chosen from a uniform distribution on the interval $[-1,1]$ and $\alpha$ is the amplitude of the vertical noise. After time $t$ crosses $t_i$ the form of $f_{\textrm{z},i}^{\textrm{n}}$ gets modified to:
\begin{equation}
\label{eq-13}
f_{\textrm{z},i}^{\textrm{n}}(t)=\frac{\alpha-\alpha\sqrt{1+4\textrm{tan}\theta+4\textrm{tan}^2\theta-8\xi_{\textrm{z},i}(t)\textrm{tan}\theta}}{2\textrm{tan}\theta}
\end{equation}
where
\begin{equation*}
\textrm{tan}\theta=\frac{1}{2}\left\{1+\textrm{exp}\left(-\frac{t-t_i}{\tau}\right)\right\}.
\end{equation*}

The above form ensures that after $t_i$ the noise becomes continually biased towards the downward direction (see figure \ref{fig:fig2sub}). The characteristic time scale for this biasing is given by $\tau$.

For setting the units of measurement we choose $R=1$ and $\Delta t=1$. The summary of the parameters,
and their typical values, as used in our simulations are provided in table~\ref{tab:tab-1}. As initial condition the birds are assigned randomly oriented velocities and are uniformly distributed in a cylinder of diameter $D$ and of height $\Delta h$.

\begin{table}
\begin{center}
\caption{The parameters with their brief explanation and values as we most often used them in our
simulations. We find that the most relevant parameters to the problem of landing are $\alpha$, $C$, $\sigma$ and $N$. 
The values of the parameters were chosen so to obtain a biologically relevant landing scenario. For setting the units of measurement we choose $R=1$ and $\Delta t=1$, which makes the horizontal interaction radius (the radius of $\mathcal{N}$) 1.
\label{tab:tab-1}}
\end{center}
\begin{center}

\begin{tabular}{|c|c|c|c|} \hline

   Param. &  Description &  Value & Dimension \\ \hline
    $\alpha$ & the coefficient of the vertical noise & $0.2095$ &�$R/\Delta t$ \\
    $C$ & the coefficient of $f_\textrm{z}^\textrm{h}$ & $0.81$ &�$R/\Delta t$ \\
    $\sigma$ & the standard deviation of the distribution of $t_i$ & $5000$ & $\Delta t$\\
    $N$ & the number of birds & $300$ & -\\
    $v$ & the speed of the birds &  $0.1$ &�$R/\Delta t$ \\
    $D$ & the horizontal diameter of the flock & $20$ &�$R$ \\
    $d$ & the ``size'' of the bird & $1/6$ &�$R$ \\
    $h$ & the optimal height of flight & $100$ & $R$ \\
    $\Delta h$ & the effective width of the flock & $2$ & $R$ \\
    $\tau$ & the timescale of energy depletion & $5000$ & $1/\Delta t$ \\
    $\mu$ & the mean of the distribution of $t_i$ & $50000$ & $\Delta t$ \\
    $A$ & the coefficient of $f_\textrm{z}^\textrm{r}$ and $\mathbf{f}_\textrm{xy}^\textrm{r}$ & $100$ & $1/\Delta t$ \\
    $B$ & the coefficient of $\mathbf{f}_\textrm{xy}^\textrm{c}$ & $100/3$ & $1/\Delta t$ \\
    $\beta$ & the coefficient of the horizontal noise & $0.05$ & $R/\Delta t$ \\ \hline
    
\end{tabular}
\end{center}
\end{table}

For a rough fitting of the model to real flocks we took data from \cite{data}. Based upon their measurements it seems reasonable to take $v\approx10~\text{m}/\text{s}$ and $R\approx2~\text{m}$, which makes $\Delta t\approx0.02~\text{s}$, since $v=0.1~R/\Delta t$ in our model. As a rough estimate from the simulations, we can take the time needed for the flock to land as $5000~\Delta t$. This means that our simulated flocks flying at the altitude of 200~m land in about a 100~s. Taking into account, that our model is very simplified compared to the actual dynamics of a bird flock, the different parameters of the model are in a reasonable range for real-life bird flocks. 
\subsection{Quantitative characterisation of the landing}
\label{values}

To quantitatively characterise the cohesion of the landing, we introduce a few quantities. We measure the spatial cohesion of the flock by taking the standard deviation ($\sigma_\textrm{xy}$) of the horizontal coordinates from the centre of mass as the following:
\begin{equation}
\sigma_\textrm{xy}=\sqrt{\frac{1}{N}\sum_{j=1}^N\left(\mathbf{r}_{\textrm{xy},j}-\mathbf{r}_\textrm{xy}^\textrm{\it CoM}\right)^2}
\end{equation}
where
\begin{equation*}
\mathbf{r}_\textrm{xy}^\textrm{\it CoM}=\frac{\sum^N_{j=1}\mathbf{r}_{\textrm{xy},j}}{N}.
\end{equation*}
We sample this quantity at two different times, once during flight ($\sigma_\textrm{xy}^0$) and once after the whole flock has landed ($\sigma_\textrm{xy}$), so that their ratio forms a measure to characterises the degree of coherence of the landing in space.

To characterise the temporal coherence of the landing we measured the following two quantities {\it viz.}  the standard deviation of the times at which the different birds have landed ($\sigma_{\textrm{L}}$) and the time ($T_\textrm{60}$) that elapses between the landing of the $20\%$ of the flock  and the landing of the $80\%$ of the flock. We additionally calculated the latter quantity in the case when the coupling between the birds was set to zero, i.e., when $f_{\textrm{z}}^{\textrm{a}}=0$ denoted by $T_\textrm{60}^0$. To obtain normalised measures characterising the temporal coherence we took $\sigma_{\textrm{L}}/\sigma$ and $T_\textrm{60}/T_\textrm{60}^0$.

\subsection{Ordering horizontally}

\begin{figure}
\begin{center}
\subfloat[$\beta = 0$]{\includegraphics[scale=0.2]{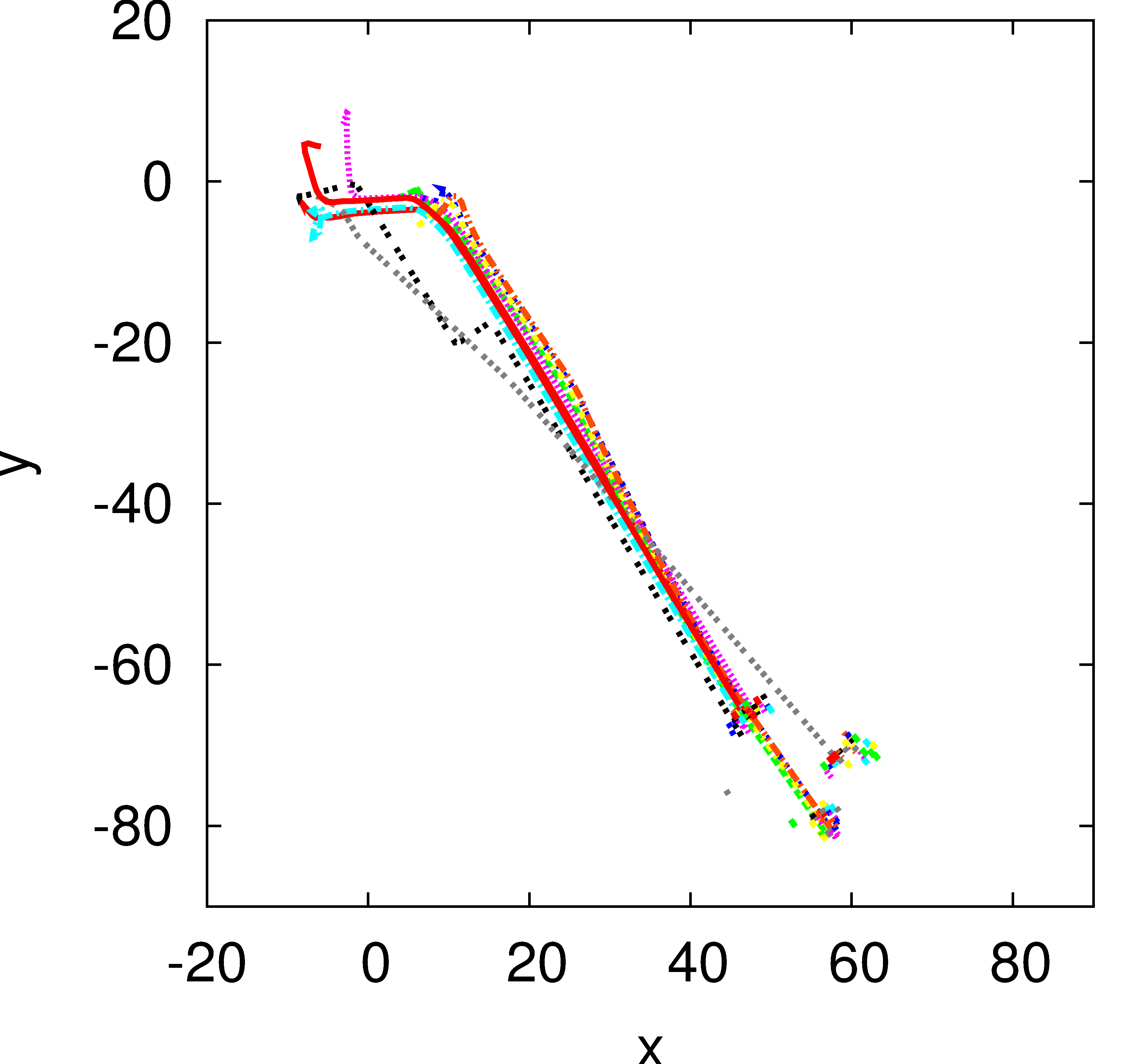}\label{fig:subfig1}}
\subfloat[$\beta = 0.03$]{\includegraphics[scale=0.2]{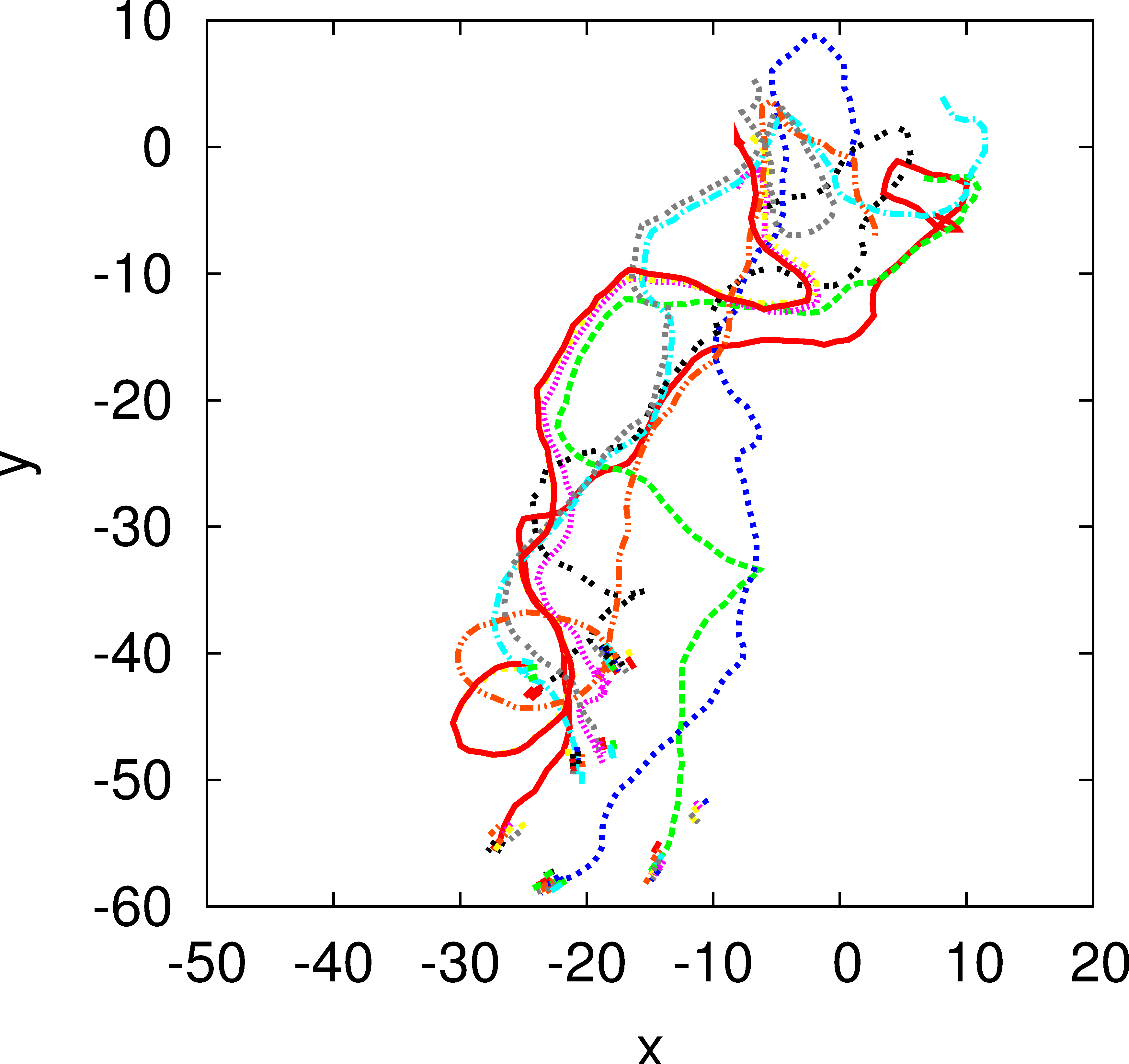}\label{fig:subfig2}}
\\
\subfloat[$\beta = 0.05$]{\includegraphics[scale=0.2]{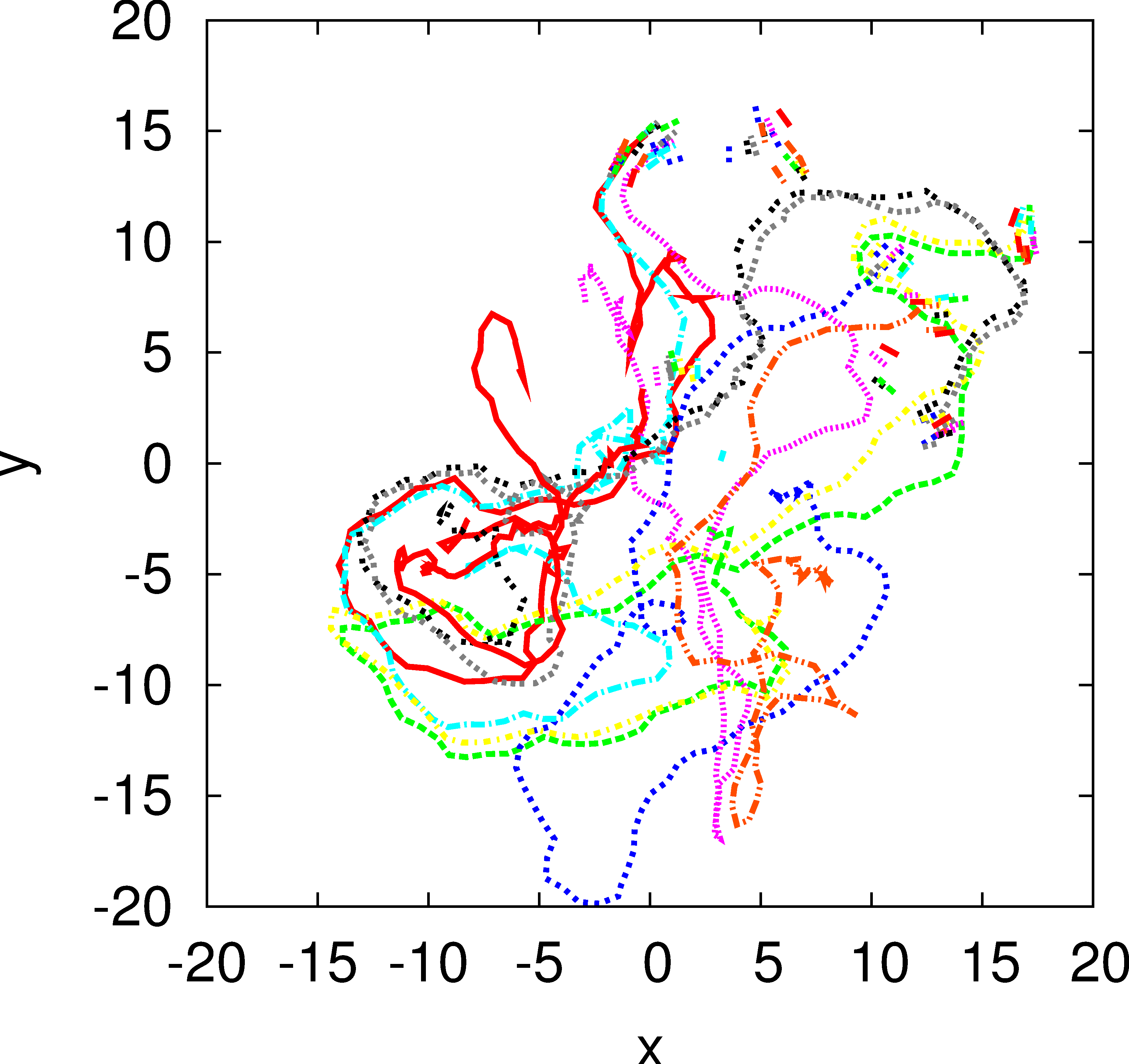}\label{fig:subfig3}}
\subfloat[$\beta = 0.07$]{\includegraphics[scale=0.2]{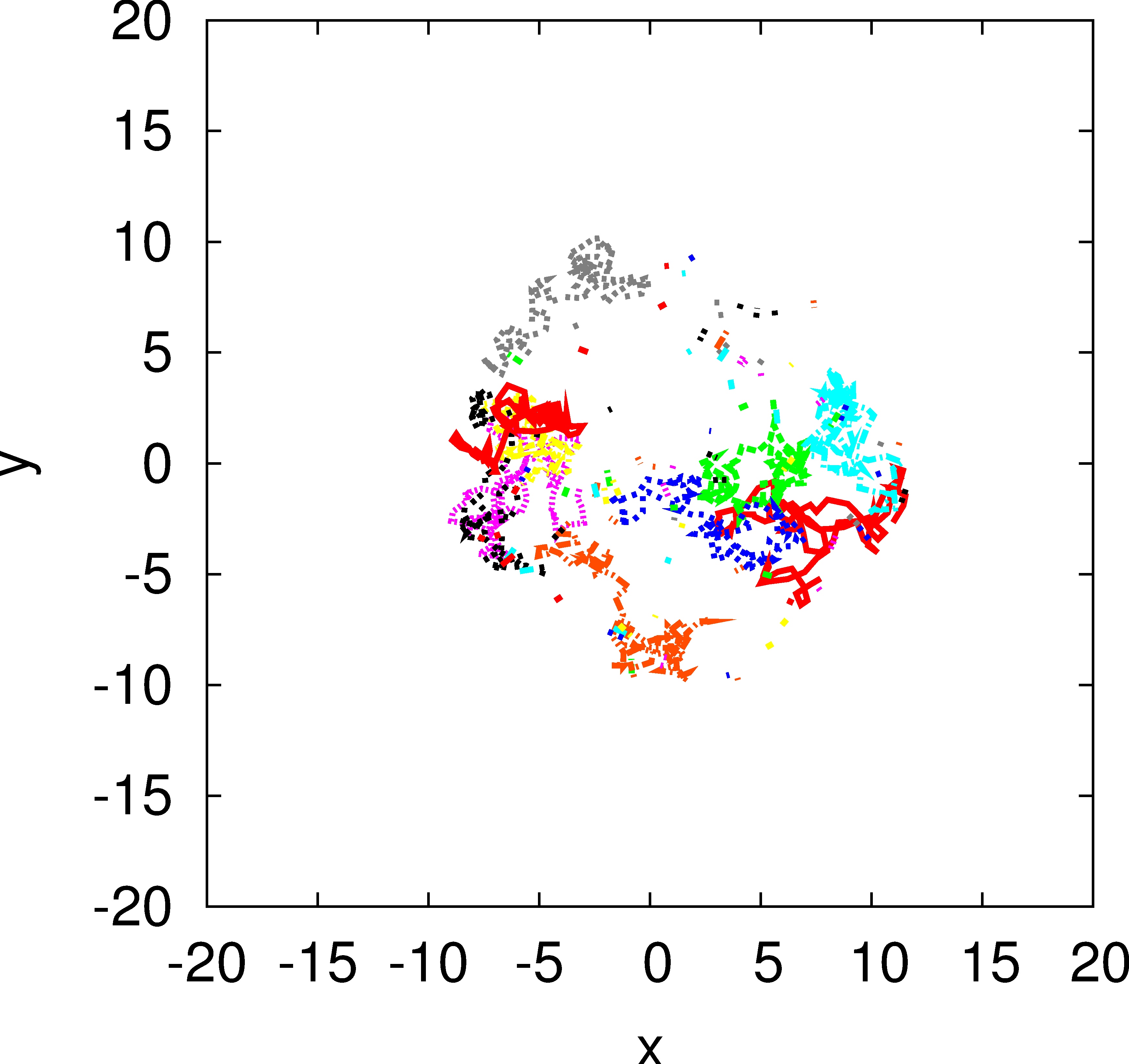}\label{fig:subfig4}}
\end{center}
\caption{The tracks of the birds projected onto the
$xy$-plane with different levels of horizontal noise. The different colours represent different birds, where 10\% of the birds were plotted with longer tracks
and the rest of the birds with just the end of their paths. From (a) to (d) we observe a transitions from an
unrealistic straight line of movement, through the biologically relevant phase to the total loss of coherent movement.
}
\label{fig-1}
\end{figure}

For obtaining a biologically relevant scenario, we adjusted the coefficient of the horizontal noise $\beta$ to an appropriate value. As can be seen in figure~\ref{fig-1}, that starting from the totally coherent, no noise situation, the increase in $\beta$ gradually destroys this coherence leading to a situation, where the centre of mass of the flock does not move at all. In this case we see the boundaries of $\mathbf{f}_\textrm{xy}^\textrm{c}$ if we plot the tracks of the birds projected unto the $xy$-plane. We chose to investigate the regime where the horizontal tracks would be similar to the one illustrated in figure~\ref{fig:subfig2}.  

\subsection{Assay of the landing}

\begin{figure}
\begin{center}
\includegraphics[width=9.0cm]{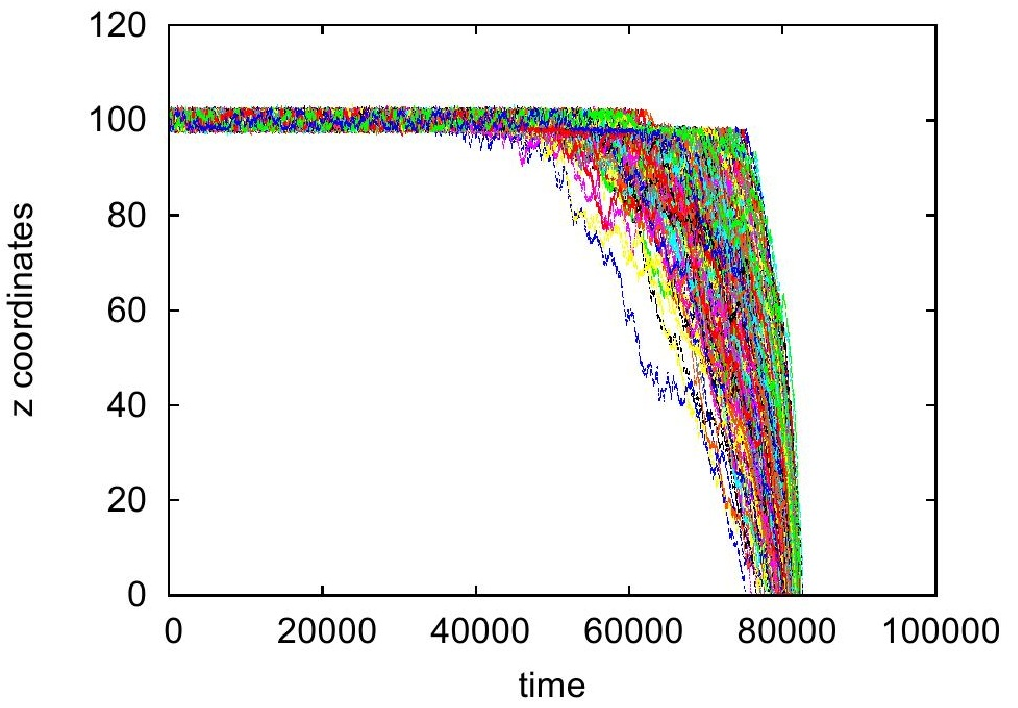}
\end{center}
\caption{The $z$-coordinates of the different birds against time. The flight around the height $h$ gives way to collective landing over time as more and more birds are increasingly biased towards moving downward. The parameters were chosen as of table~\ref{tab:tab-1}.
}
\label{fig-3}
\end{figure}

\begin{figure}
\begin{center}
\includegraphics[width=9.0cm]{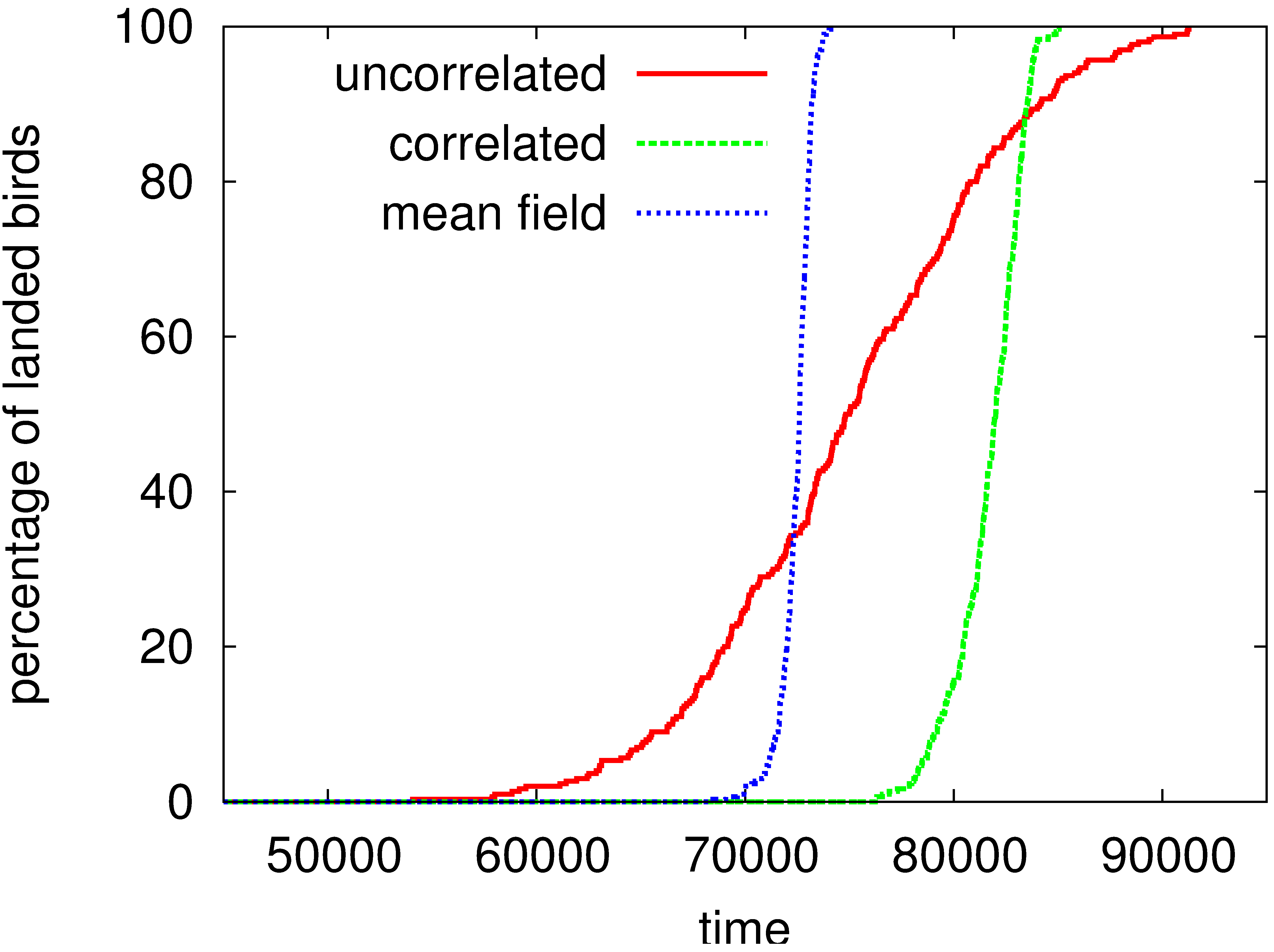}
\end{center}
\caption{The percentage of landed birds as a function of time. The red curve 
corresponds to the case when coupling between the birds is absent, i.e., $f_\textrm{z}^\textrm{a}=0$, the green one corresponds to the coupled case, while the blue curve is the mean field case, i.e., where the radius of $\mathcal{N}$ is infinity instead of $R$. It is clearly seen that in the presence of coupling, the landing is much sharper {\it viz.} the synchronisation among the birds is much greater. It is also notable, that increasing the radius of interaction to infinity does not make the landing process relevantly sharper, it merely decreases the time needed to make the decision to land. The parameters were chosen as of table~\ref{tab:tab-1}.
}
\label{fig-4}
\end{figure}

After choosing the parameters for horizontal motion we observe the time evolution of the $z$-coordinates of a similar flock in figure~\ref{fig-3}. The two states of flight, horizontal motion of the flock at a height around $h$ and the landing are clearly visible. To see whether the landings are synchronised or not we have plotted the percentage of the landed birds in the flock as a function of time, in the presence of coupling, in the absence of coupling ($f_\textrm{z}^\textrm{a}=0$) and in a mean field case (meaning the radius of $\mathcal{N}$ is infinity, not $R$) in figure~\ref{fig-4}.

As we can see, the transition in the coupled case is much sharper than in the uncoupled case, showing considerable synchronisation among the birds. It is also notable, that increasing the radius of interaction to infinity does not make the landing process relevantly sharper, it merely decreases the time needed to make the decision to land. Coherent landing arises from the interplay of three forces: $f_\textrm{z}^\textrm{h}$, the vertical noise and the averaging force. The magnitude of the bias in the noise increases with time to ultimately overcome $f_\textrm{z}^\textrm{h}$, but due to the averaging force the individual biases are, in a sense, averaged over the neighbouring birds. Thus for an individual bird it becomes ``harder'' to land when the bulk of the flock would still want to fly, and becomes ``easier'' when the latter wants to land.

\begin{table}
\begin{center}
\caption{The normalised quantities describing the coherence in space and time, with the parameter values from table~\ref{tab:tab-1}. $T_\textrm{60}^\textrm{0}$ is the value of $T_\textrm{60}$ for the zero coupling case and $\sigma_\textrm{xy}^\textrm{0}$ is the values of  $\sigma_\textrm{xy}$ during flight.}
\label{tab-2}
\end{center}
\begin{center}
\begin{tabular}{|c|c|} \hline
   Quantity &  Value \\ \hline
    ${\sigma}_\textrm{L}/\sigma$ & $0.387\pm 0.007$ \\\hline
    $T_\textrm{60}/T_\textrm{60}^\textrm{0}$ & $0.288\pm 0.007$ \\\hline
    $\sigma_\textrm{xy}^\textrm{0}/\sigma_\textrm{xy}$ & $0.639\pm 0.017$\\\hline
\end{tabular}
\end{center}
\end{table}

In table 2 we have summarised the quantitative analysis of the landing. For this we measured the quantities introduced in section \ref{values} in around hundred independent runs of the simulation, and averaged over the obtained data. Note that the normalised quantities describing temporal coherence show considerable synchronisation in the flock. We stress that this happened while a single bird had information about only a few neighbours, and not the whole flock.

\begin{figure}
\begin{center}
\includegraphics[width=9.0cm]{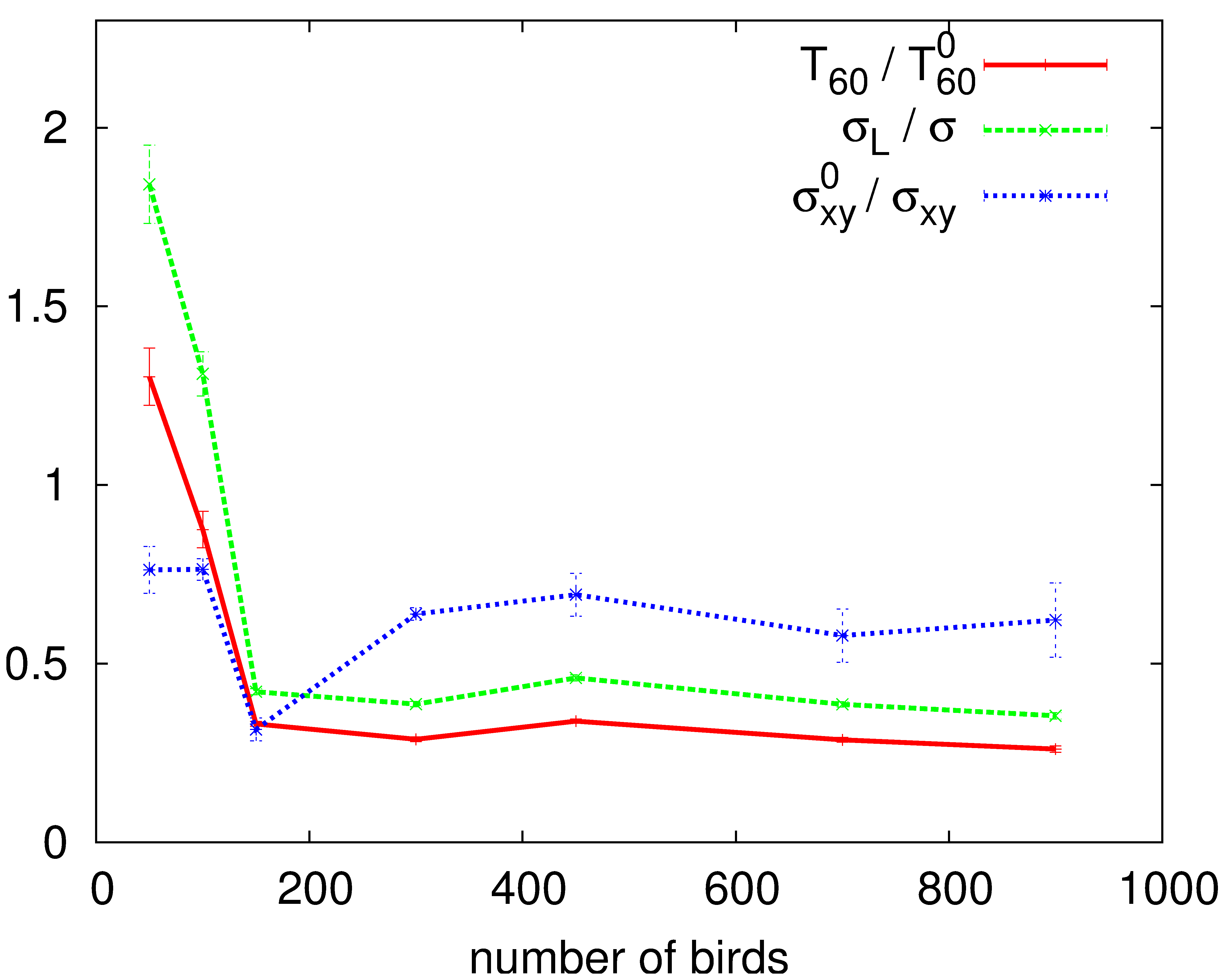}
\end{center}
\caption{The values of the most important quantities describing the collective landing for different
number of birds, while keeping the density of the birds constant with the changing of diameter $D$ (all other parameters are as of table 1). Note that $T_\textrm{60}/T_\textrm{60}^\textrm{0}$ is well below unity and this property persists for much
larger flocks than the one used in most of our measurements.
}
\label{fig-5}
\end{figure}

\begin{figure}
\begin{center}
\includegraphics[width=9.0cm]{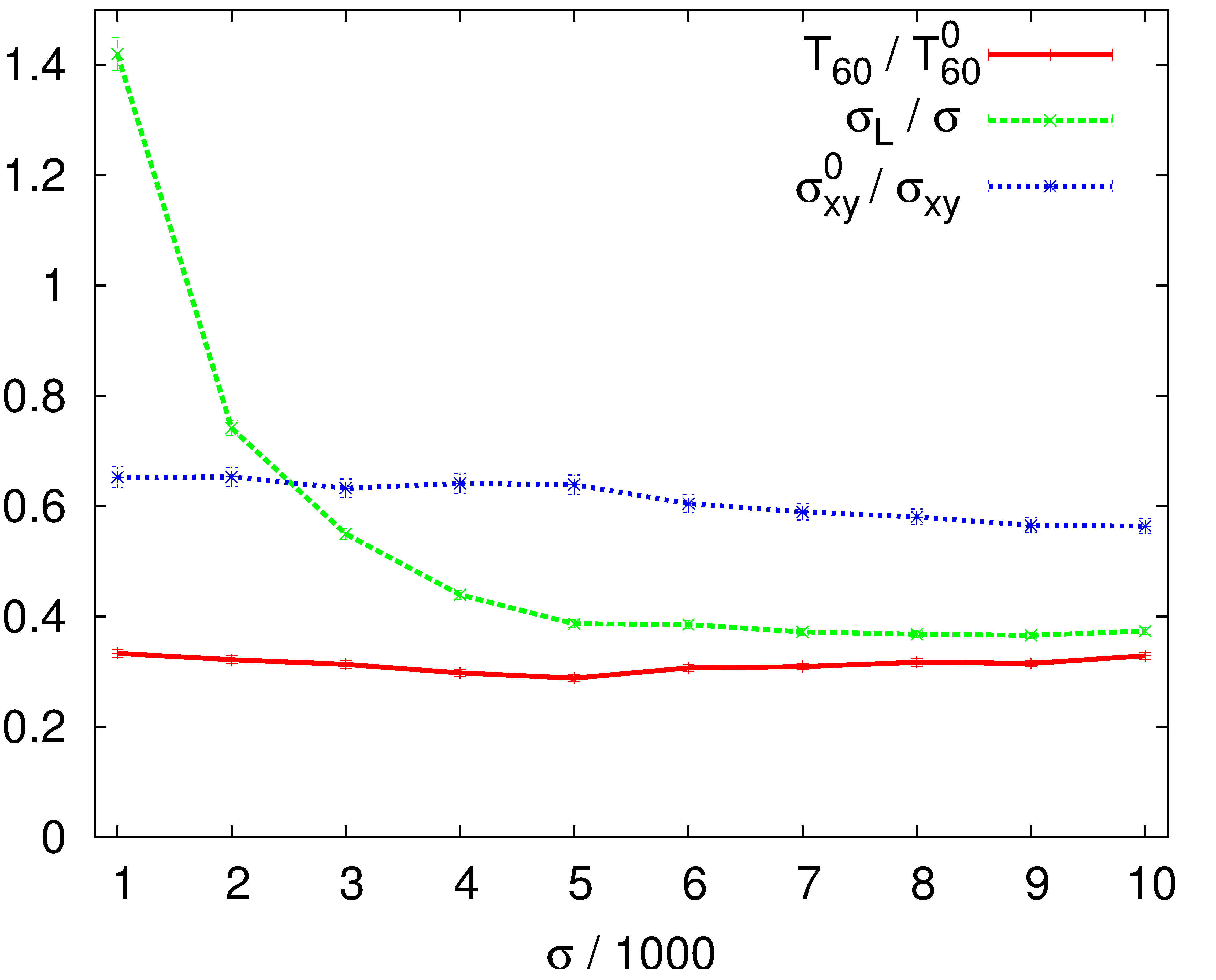}
\end{center}
\caption{The values of the most important quantities describing the collective landing for different
standard deviation of the depletion times (all other parameters are as of table 1). Note that $T_\textrm{60}/T_\textrm{60}^\textrm{0}$ is well below unity. For small values of $\sigma$, the ratio $\sigma_\textrm{L}/\sigma$ grows considerably due to the fact, that factors such as the size
of the birds and the height of the flock, become important in deciding the time taken by the whole flock to land.
}
\label{fig-6}
\end{figure}

\begin{figure}
\begin{center}
\subfloat[]{\includegraphics[scale=0.27, totalheight = 5 cm]{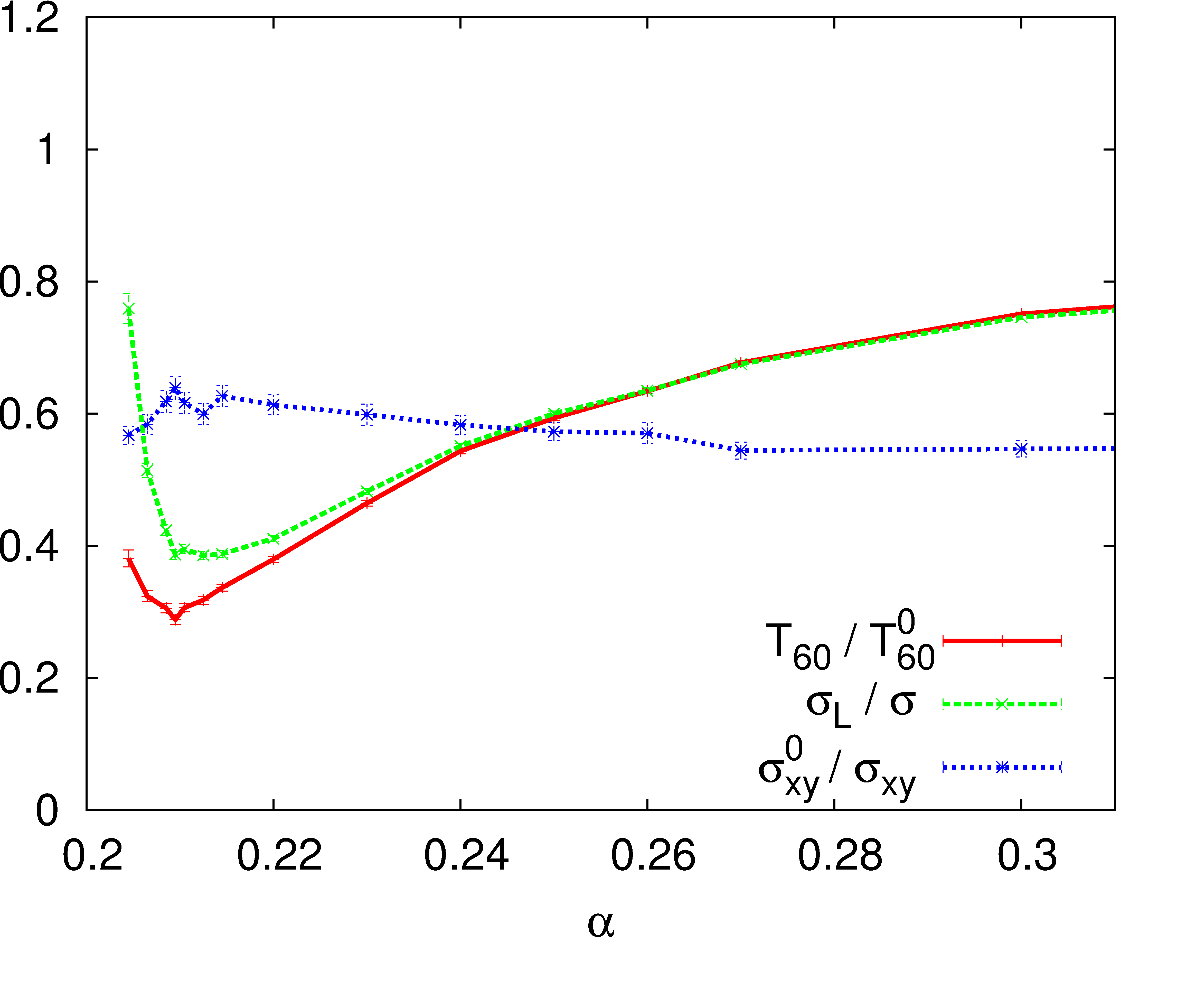}\label{fig:fig7a}}
\subfloat[]{\includegraphics[scale=0.27, totalheight = 5 cm]{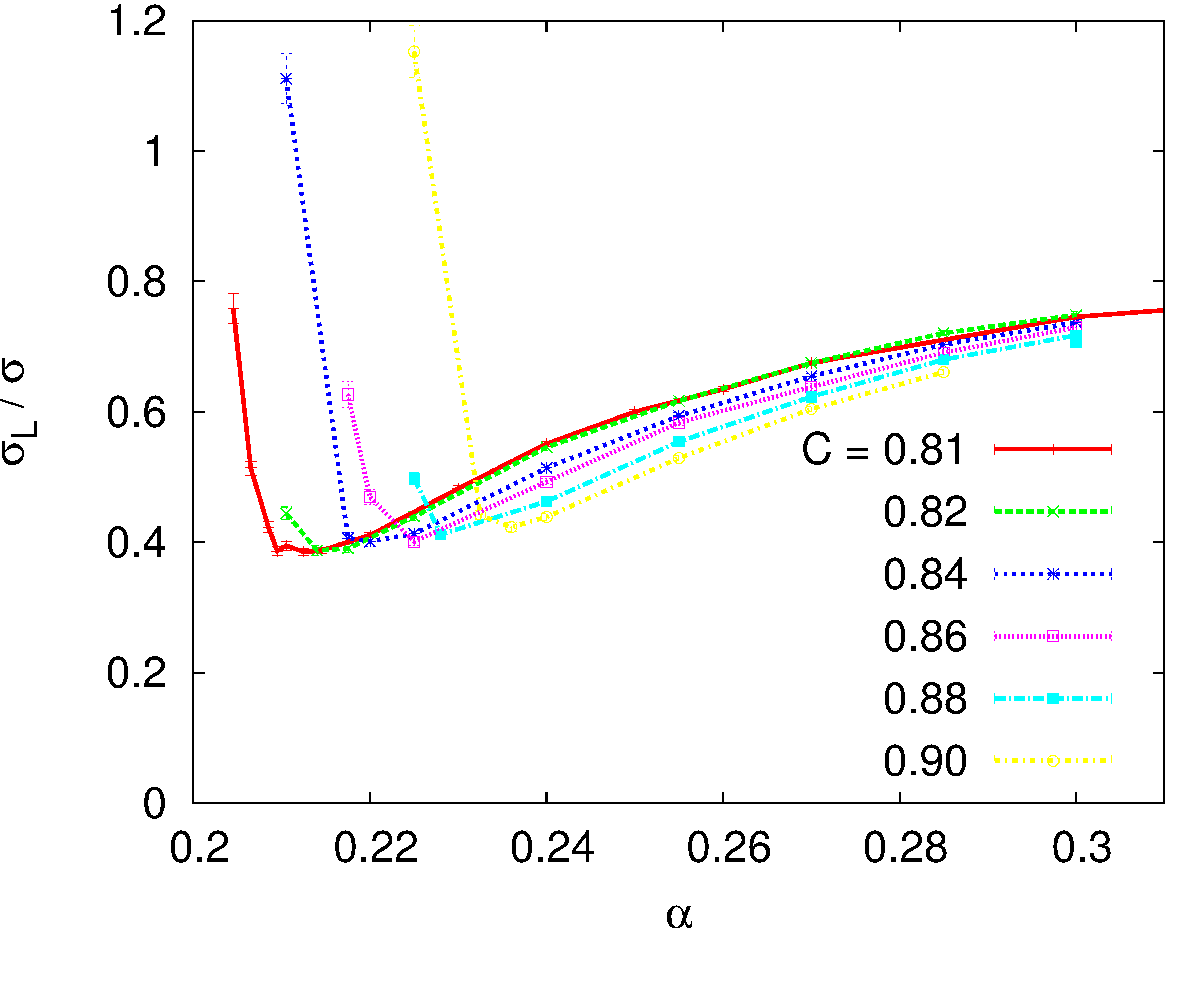}\label{fig:fig7b}}
\end{center}
\caption{
(a) The values of the most important quantities describing the collective landing for different magnitudes of the vertical noise (all other parameters are as of table~\ref{tab:tab-1}). For values of $\alpha$ slightly less than the smallest shown on the graph no landing occurs. (b) The temporal coherence of the flock as a function of the magnitude of the vertical noise and $f_{\textrm{z}}^\textrm{h}$ (all other parameters are as of table~\ref{tab:tab-1}). The place of the minimum in $\alpha$ shifts to the right, and the value of the minimum increases as $f_{\textrm{z}}^\textrm{h}$ becomes stronger. For values of $\alpha$ slightly less than the smallest shown on the graph no landing occurs.
}
\label{fig-7}
\end{figure}

In figures \ref{fig-5}, \ref{fig-6} and \ref{fig-7} we plot the quantities mentioned in table~\ref{tab-2}, as functions of the number of birds in the flock, the
standard deviation of the distribution of $t_i$-s, and the magnitude of the vertical noise, respectively. While changing the number of birds in the flock, we keep the density of the birds constant by appropriately changing the horizontal diameter $D$ of the flock. Note that our previous statements about temporal coherence hold for considerable changes of these attributes of the flock.

In figure~\ref{fig-5} we see that decreasing the number of birds, while keeping the ratio of the number of birds and the volume set by the boundaries of the forces $f_\textrm{z}^\textrm{h}$ and $\mathbf{f}_\textrm{xy}^\textrm{c}$ a constant, makes the flock less coherent. This is due to the fact that with the decrease in the number of birds, the fluctuations in the number of neighbours, with respect to the individual birds, become important. In figure~\ref{fig-6} we see that for small $\sigma$-s the value of ${\sigma}_\textrm{L}/\sigma$ grows well above unity. This is because there is an inherent difference in landing times, due to the birds starting from different heights and thus arriving at the ground at different times, and also due to the fact that repulsion prevents the birds from landing on top of each other. The effect of this inherent difference becomes noticeable compared to low $\sigma$ values.

Figure~\ref{fig-7} shows that for any given coefficient of $f_\textrm{z}^\textrm{h}$ and $C$, there is maximum of temporal coherence or, alternatively a minima of $T_\textrm{60}/T_\textrm{60}^\textrm{0}$ and ${\sigma}_\textrm{L}/\sigma$, as a function of the noise. The interplay between $f_\textrm{z}^\textrm{h}$ and the vertical noise becomes apparent from figure~\ref{fig:fig7b}. The location of the minima as a function of $\alpha$ (the coefficient of the vertical noise) shifts to right, and its value increases (in ${\sigma}_\textrm{L}/\sigma$)  with the increase of the coefficient of $f_\textrm{z}^\textrm{h}$. To explain this, let us increase $\alpha$ from zero as we keep $C$ a constant. At small values of $\alpha$ we see that there is no landing, as the noise is not strong enough to overcome $f_\textrm{z}^\textrm{h}$. When $\alpha$ is larger (right side of figure~\ref{fig:fig7b}) the temporal coherence of the flock decreases with the increase of $\alpha$, as one would intuitively think. Between these two there is regime with a non-trivial minimum of ${\sigma}_\textrm{L}/\sigma$ (maximum of temporal coherence), where the increase in the magnitude of the noise actually increases temporal coherence within the flock (see e.g. noise-induced ordering \cite{noiseinduced}, stochastic resonance \cite{stochasticorder}). The increase of $C$ decreases the maximum of temporal coherence the flock can reach and increases the magnitude of noise needed to reach this maximum.

Our model, naturally, uses assumptions which are results of simplifications
of the true complexity of a flock of birds. We assume local interactions (that these
dominate the landing process) in addition to the a priori assigned intent of the birds.
A possibility would be to include the reaction of a single bird to the behaviour of the whole flock
(taking this into account would need a number of further arbitrary assumptions), but
we have restricted the model to mostly local interactions, with one bird interacting with an estimated 8-12 birds on average.

\section{Conclusion}

In this report, we have investigated a model where the concepts of collective decision making and
collective motion are intertwined. In particular, we have introduced a simple phenomenological
model for the landing of a flock of birds. In the model birds are only influenced by the dynamical
variables, like position and velocity, of other birds in their immediate neighbourhood. Heterogeneity
is introduced in the flock in terms of a depletion time after which a bird feels an increasing bias to
move towards the landing surface. 
The stochastic nature of the bias ensures that the external and internal effects influencing the flock are also included.
Through our model we have demonstrated a mechanism by which animals in large group can arrive
at an egalitarian decision about the time of switching from one activity to another in the absence of a
leader. Our results suggest that the coherence of the
collective action of landing can be enhanced by the random perturbations. 

\section*{Acknowledgement}
This research was supported by the EU ERC COLLMOT project.

\end{document}